\documentclass[sigconf]{acmart}
\AtBeginDocument{%
  }

\setcopyright{acmlicensed}
\copyrightyear{2018}
\acmYear{2018}
\acmDOI{XXXXXXX.XXXXXXX}
\acmConference[Conference acronym 'XX]{Make sure to enter the correct
  conference title from your rights confirmation email}{June 03--05,
  2018}{Woodstock, NY}
\acmISBN{978-1-4503-XXXX-X/2018/06}

\acmISBN{}
\acmDOI{}

\usepackage{xspace}
\usepackage{multirow} 
\usepackage{tikz}

\newcommand{\rbIV}{Robust `04\xspace}
\newcommand{\dlXIX}{DL `19\xspace}
\newcommand{\dlXX}{DL  `20\xspace}

\usepackage[normalem]{ulem}
\usepackage{mdframed}
\usepackage{balance}
\usepackage{subcaption}
\usepackage{float}
\usepackage{cellspace} 

\begin{document}

\title{Embedding Surgery: Localized Updates for Adaptive Ranking Correction in Dense Retrieval}
\titlenote{Accepted at ACM CIKM 2026. The definitive Version of Record is available at \url{https://doi.org/10.1145/3799682.3840777}.}


\author{Maddalena Amendola}
\email{maddalena.amendola@iit.cnr.it}
\affiliation{%
  \institution{IIT-CNR}
  \city{Pisa}
  \country{Italy}
}

\author{Antonio Mallia}
\email{antonio@seltz.ai}
\affiliation{%
  \institution{Seltz}
  \city{San Francisco}
  \state{California}
  \country{USA}
}

\author{Raffaele Perego}
\email{raffaele.perego@isti.cnr.it}
\affiliation{%
  \institution{ISTI-CNR}
  \city{Pisa}
  \country{Italy}
}

\renewcommand{\shortauthors}{Amendola et al.}


\begin{abstract}
Dense retrieval systems are core components of modern search engines, recommendation platforms, and retrieval-augmented generation pipelines. They encode documents and queries into dense embeddings, enabling efficient semantic search via vector similarity. However, because document embeddings are computed offline and stored in static indexes, these systems struggle to adapt to user feedback or evolving search intent.
To address this limitation, we introduce \emph{embedding surgery}, a lightweight approach for adaptive ranking correction in dense retrieval. The method applies localized, minimal updates to selected document embeddings at query time, guided by editorial feedback, user interactions, or pseudo-labels from large language models. We formulate embedding surgery as a convex optimization problem that enforces ranking constraints while minimizing modifications to the affected document representations.
We integrate embedding surgery into standard dense retrieval pipelines and evaluate it on TREC Deep Learning, TREC Robust, TREC CAsT, and MS MARCO benchmarks. Results show consistent improvements (e.g., up to +60.64\% relative improvement in nDCG@10 on DL-Hard under editorial feedback), even under noisy or shifting feedback, with low computational cost and without disrupting the global structure of the embedding space. Extensive experiments show that ranking corrections propagate to semantically related queries and that embedding updates can be applied safely and efficiently to scalable Approximate Nearest Neighbor indexes via simple in-place overwriting, without requiring costly index reconstruction. Finally, embedding surgery complements query adaptation methods such as CoRocchio, yielding additional gains while being more robust to noisy feedback.
\end{abstract}


\received{20 February 2007}
\received[revised]{12 March 2009}
\received[accepted]{5 June 2009}

\maketitle

\section{Introduction}

Vector search systems have become the backbone of modern search engines, recommendation systems, retrieval-augmented generation (RAG), and Q\&A applications. These systems encode documents and queries as embeddings in a learned latent space and retrieve candidates via vector similarity search, typically implemented using Approximate Nearest Neighbor (ANN) methods. In practice, embeddings are produced by neural bi-encoders and deployed in a pipeline where document representations are computed and indexed offline, while queries are encoded at runtime. While this design enables scalability and efficient retrieval, it also results in static document embeddings, making it difficult to incorporate user feedback or adapt to ranking errors.

In large-scale Information Retrieval (IR) systems, this rigidity is a significant limitation, particularly as search topics evolve or specific ranking errors emerge. Prior work has explored several strategies to mitigate this issue, including online learning-to-rank methods \cite{10.1145/2835776.2835804,10.1145/3269206.3271686,10.1007/978-3-030-45439-5_28} and hybrid pipelines that combine static embeddings with dynamic signals or user interaction feedback \cite{10.1145/3471158.3472233,10.1007/978-3-031-28244-7_5,10.1145/3631939}. Another common approach is to fine-tune document encoders on batches of error cases, followed by periodic re-indexing. However, model fine-tuning and re-indexing are computationally expensive and unsuitable for the timely correction of specific ranking errors (e.g., biases affecting individual queries).

A more flexible approach is needed to efficiently adapt document embeddings, allowing retrieval systems to incorporate feedback and perform targeted ranking corrections. This approach allows static retrievers to address semantic mismatches and refine rankings based on explicit or implicit feedback signals, such as editorial relevance annotations or user interactions on the search engine results page (SERP). These signals can reveal misranked results and guide localized embedding updates to correct the ranking order. When direct user feedback is limited, auxiliary supervision from large language models (LLMs) can also help identify misalignment between query intent and document content.

In this paper, we introduce \textbf{embedding surgery}, a paradigm for flexible, interactive, and high-precision retrieval based on lightweight updates to stored document embeddings. Rather than retraining models or rebuilding indexes, embedding surgery directly modifies the embeddings of retrieved documents to correct query-specific ranking errors identified through feedback. These localized modifications enable adaptive retrieval with low computational overhead while preserving the overall geometry of the embedding space.
We consider three sources of supervision to guide these updates: (i) editorial judgments from expert assessors, which provide high-quality signals for correcting known ranking errors, enforcing business rules, or addressing fairness and bias concerns; (ii) user interaction signals (e.g., clicks and dwell time), simulated via counterfactual click models in the absence of real query logs; and (iii) auxiliary supervision from LLMs, which can approximate relevance judgments and help identify ranking errors~\cite{qin-etal-2024-large,10.1145/3726302.3730159,10.1145/3578337.3605136}. However, the use of LLMs as assessors should be approached with caution: \citet{DBLP:conf/evia/0001D25} argue that LLM-generated labels exhibit systematic weaknesses that limit their ability to fully replace human relevance assessments.

Our goal is to show that local, feedback-driven updates to document embeddings can enforce ranking constraints while preserving the global structure of the embedding space. We formulate embedding surgery as a constrained optimization problem in which embeddings of selected retrieved documents are adjusted to satisfy ranking constraints derived from supervision signals while remaining close to their original representations. The resulting optimization problem is a convex quadratic program. For a single pairwise constraint, the optimal update admits a closed-form solution; for batches of constraints, the corresponding quadratic program can be solved efficiently using standard optimization solvers.

We evaluate our approach on multiple standard benchmarks, including TREC Deep Learning (DL) 2019–2020 and Hard, TREC \rbIV, TREC CAsT 2019, and MS MARCO. Results show that embedding surgery can precisely correct ranking errors with minimal computational cost, yielding consistent improvements of up to +60.64\% relative nDCG@10. The method remains effective under noisy feedback, propagates ranking corrections to semantically related queries, and complements query adaptation approaches such as CoRocchio \cite{ZhuangLiZuccon2022}, yielding additional gains. Furthermore, large-scale experiments demonstrate that these embedding updates preserve the global structure of the embedding space and can be applied safely and efficiently via in-place overwriting in graph-based indexes such as HNSW and cluster-based indexes such as IVF.

The main contributions of this work are:
\begin{itemize}
    \item A formalization of embedding surgery as a constrained optimization problem, enabling query-time correction of ranking errors while keeping embedding updates local and minimal;
    \item A unified framework for integrating embedding surgery with editorial, click-based, and LLM-derived feedback in dense retrieval pipelines;
    \item An extensive evaluation across four dense retrieval models and seven IR benchmarks, demonstrating consistent improvements, positive transfer to semantically related queries, robustness to noisy feedback, adaptability across heterogeneous feedback sources, effectiveness in out-of-domain settings, and complementarity with query adaptation methods;
    \item A large-scale analysis over 56k queries on ANN index structures (IVF and HNSW), showing that updates remain localized, preserve retrieval effectiveness, and can be applied safely and efficiently without costly index reconstruction.
\end{itemize}
Together, these results position embedding surgery as a practical approach for adaptive, feedback-driven ranking correction in dense retrieval systems. 

\section{Related work}
\label{sec:related}
Our embedding surgery approach introduces a novel class of techniques for post-hoc correction of ranking errors in dense retrieval, based on constrained and minimal modifications to document embeddings. To the best of our knowledge, this direction has not been explored in prior work.

Existing literature has instead focused on three complementary and largely orthogonal lines of research: (1) online learning-to-rank (OLTR) methods that exploit user interactions to adapt the ranking function; (2) strategies for determining when and how dense retrievers should be updated, particularly under distribution shift or evolving corpora; and (3) approaches that leverage user interaction signals to adapt query representations to improve retrieval effectiveness.
In the following, we survey these related areas:

\paragraph{\textbf{Online learning to rank.}}
Online Learning to Rank (OLTR) aims to optimize ranking models through continuous real-time user interactions, balancing exploration of new rankings with exploitation of observed feedback. It is particularly suited to scenarios where historical data are scarce or where the retrieval context evolves rapidly, requiring models to be initialized or adapted online. The seminal work by \citet{10.1145/1553374.1553527} formulated OLTR as an online dueling bandit problem, showing how user feedback can drive iterative improvements without relying on offline supervision. Subsequent work has focused on improving efficiency and scalability. \citet{10.1145/2835776.2835804} proposed Multileave Gradient Descent to accelerate OLTR by evaluating multiple rankers simultaneously, while \citet{10.1145/3269206.3271686} introduced differentiable, unbiased OLTR methods that extend the approach to neural rankers and enable the optimization of non-linear models. Similarly to our approach, counterfactual modeling can be incorporated to obtain unbiased updates from user interactions \cite{10.1007/978-3-030-45439-5_28}.

\textit{Unlike OLTR methods, which update ranking functions through continuous interaction, embedding surgery assumes a well-trained retriever and targets residual, query-specific ranking errors that inevitably arise in practice. It operates directly on document embeddings to perform localized corrections, enabling immediate adaptation without exploration or sustained user interaction.}

\paragraph{\textbf{Updating/retraining retrievers.}}
Another line of work investigates when and how dense retrievers should be updated to remain effective under distribution shifts or evolving corpora. The key challenge is balancing adaptability with efficiency. Unlike many NLP tasks, retrieval requires maintaining large embedding indexes, which can become stale as the retriever evolves. Fine-tuning on new data introduces embedding drift, potentially degrading performance on previously indexed content. Since recomputing embeddings after each update is computationally expensive, continual adaptation remains challenging. Prior work shows that catastrophic forgetting also occurs in dense retrieval \cite{catastrophic-forgetting}, and that continual learning strategies can help mitigate it.
\citet{DBLP:conf/acl/KoKKKLSLK25} introduced GradNormIR, an unsupervised method for detecting whether a new corpus is out-of-distribution for a given retriever by analyzing gradient-norm perturbations of document embeddings, enabling proactive updates. Complementarily, \citet{10.1007/978-3-031-88708-6_6} proposed MURR, a strategy that fine-tunes retrievers on new data while preserving backward compatibility with existing embeddings. By leveraging regularized replay of past training signals, MURR avoids costly re-encoding of the entire collection, enabling more efficient incremental maintenance. 

\textit{These approaches are orthogonal to ours and primarily address long-term retriever maintenance under distribution shift. In contrast, embedding surgery focuses on short-term, query-specific error correction through localized updates without retraining or re-indexing.}

\paragraph{\textbf{Improving query representations.}}
Another line of work improves retrieval effectiveness by adapting query representations using user interaction signals derived from counterfactual modeling. CoRocchio \cite{ZhuangLiZuccon2022} extends the classical Rocchio algorithm~\cite{Rocchio1971} to dense retrieval by updating query embeddings as a combination of the original query and embeddings of clicked documents, with weights adjusted by click frequency and debiased for position effects.
In a similar setting, CoDIME \cite{FFPT2025} estimates the importance of each query embedding dimension based on its alignment with document representations identified from debiased clicks. Dimensions strongly aligned with frequently clicked documents are preserved, while misaligned or noisy dimensions are suppressed. Both CoRocchio and CoDIME exemplify query-side adaptation, in contrast to embedding surgery, which performs document-side, query-specific corrections to ranking behavior. 

\textit{Query adaptation approaches are complementary to embedding surgery: in Section \ref{sec:corocchio} we show that combining our approach with CoRocchio yields additional gains over either method alone.}

\section{Embedding Surgery}
\label{sec:methodology} 

Dense retrieval is a vector-based IR paradigm in which queries and documents are independently encoded into a shared latent space using a bi-encoder architecture. The encoders are trained jointly with contrastive learning, and similarity between embeddings is typically computed using the inner product or cosine similarity. For scalability, document embeddings are precomputed offline and indexed using Approximate Nearest Neighbor (ANN) search systems such as FAISS \cite{douze2025faisslibrary,DBLP:journals/corr/JohnsonDJ17}, Falcon \cite{10.14778/3749646.3749655}, ScaNN \cite{10.5555/3524938.3525302}, HNSW \cite{8594636}, MRNG \cite{10.14778/3303753.3303754}, and others \cite{10.1007/s00778-024-00864-x}. This design enables efficient top-\textit{k} retrieval while alleviating the vocabulary mismatch issues of lexical methods.

Formally, let $\mathbf{q} \in \mathbb{R}^d$ denote the embedding of a query $q$, and $\mathbf{d}_i \in \mathbb{R}^d$ the embedding of a document $d_i \in \mathcal{D}$. The relevance score is computed as:
\[
\text{score}(q, d_i) = \langle \mathbf{q}, \mathbf{d}_i \rangle ,
\]
where $\langle \cdot, \cdot \rangle$ denotes the similarity function. The top-\textit{k} documents with the highest scores, denoted $\pi = [d_1, d_2, \ldots, d_k]$, are returned as the result set.

The goal of embedding surgery is to apply small, localized modifications to document embeddings so that the resulting ranking better reflects feedback signals, while preserving the global structure of the embedding space. 
Formally, given a query $q$ with embedding $\mathbf{q} \in \mathbb{R}^d$ and a ranked list of retrieved documents $\pi = [d_1, d_2, \ldots, d_k]$, where each document $d_i$ is represented by an embedding $\mathbf{d}_i$, we assume access to feedback signals $\mathcal{R}$ that specify desired corrections to the ranking. Each element of $\mathcal{R}$ is a pairwise preference $(d_r, d_n)$, where both $d_r$ and $d_n$ belong to the retrieved set $\pi$, indicating that $d_r$ should be ranked above $d_n$ for query $q$. Embedding surgery aims to compute updates $\Delta \mathbf{d}_i$ to the embeddings of the retrieved documents that enforce these constraints while minimizing the overall perturbation to the latent space.

The optimization problem is formally defined as:
\begin{equation}
    \min
    \quad \sum_{i=1}^{k} \|\Delta \mathbf{d}_i\|_2^2 
    \label{eq:surgery}
\end{equation}
subject to \(\quad \langle \mathbf{q}, \mathbf{d}_r + \Delta \mathbf{d}_r \rangle \geq \langle \mathbf{q}, \mathbf{d}_n + \Delta \mathbf{d}_n \rangle + \epsilon, \quad \forall (d_r, d_n) \in \mathcal{R}\), where:
\begin{itemize}
\item $\Delta \mathbf{d}_i$ is the correction applied to the original embedding $\mathbf{d}_i$;
\item $\| \cdot \|_2$ denotes the L2 norm, penalizing large deviations from the original vectors;
\item $\epsilon > 0$ defines a margin that enforces separation between the relevance scores for query $q$ of documents $d_r$ and $d_n$.
\end{itemize}

The objective function encourages minimal movement in the embedding space, ensuring that the adjusted vectors remain close to their original positions. At the same time, the constraints guarantee that the resulting ranking is consistent with the feedback. Because the optimization problem is convex and quadratic, it admits a unique global optimum and can be solved efficiently using standard solvers such as CVXPY. 
The corrected embeddings are then obtained as:
\[
\mathbf{d}_i' = \mathbf{d}_i + \Delta \mathbf{d}_i.
\]

The general formulation allows for several variants, depending on which embeddings of each pair are permitted to change. These variants can emphasize different correction behaviors:

\begin{enumerate}
	\item \textbf{Symmetric (default)}: Both embeddings are allowed to adjust, resulting in a balanced correction that minimizes total movement while satisfying the feedback constraint. 
	\item \textbf{Demotion}: The embedding of the more relevant document is fixed ($\Delta \mathbf{d}_r = \mathbf{0}$), and only the less relevant one is modified. This approach demotes the less relevant document in the ranking while leaving the relevant document unchanged.
	\item \textbf{Promotion}: The embedding of the less relevant document is fixed ($\Delta \mathbf{d}_n = \mathbf{0}$), and only the more relevant one is modified. This approach corresponds to promoting the more relevant document upward in the ranking.
\end{enumerate}

\subsection{Feedback Models for Surgery Triggers}
\label{sec:feedback}

Embedding surgery relies on external signals to identify when ranked results for a query are incorrect or misaligned with user or system expectations. Such signals may come from editorial annotations (e.g., manually curated relevance judgments, fairness assessments, or business-driven constraints), user interactions with SERPs, or auxiliary re-ranking signals obtained, for example, from trusted LLMs. 
In this work, we consider three representative (but not exhaustive) sources of relevance feedback: \textit{editorial}, \textit{click-based}, and \textit{LLM-based}.

\paragraph{\textbf{Editorial and LLM-based Feedback.}}
\label{sec:LLM-based}

Editorial feedback consists of human relevance judgments that establish the gold standard for ranking. These assessments, typically provided by expert annotators, represent the most reliable form of supervision and are commonly used as an oracle for evaluating ranking model performance. In this paper, we leverage editorial feedback to assess embedding surgery under ideal conditions and to analyze how the resulting modifications reshape the latent space.

Since such feedback is costly and time-consuming to obtain, we also investigate whether LLMs can provide a practical alternative. Using a state-of-the-art LLM $\mathcal{L}$, we estimate the alignment between query intent and the content of retrieved documents, enabling the identification of potential ranking errors in settings where explicit or implicit user feedback is sparse or unavailable.

For both feedback sources, let $\pi = [\pi_1, \pi_2, \dots, \pi_k]$ denote the ranked list produced by the dense retrieval model for a query $q$, and let $\sigma = [\sigma_1, \sigma_2, \dots, \sigma_k]$ represent the alternative ranking generated by the selected feedback source (Editorial or $\mathcal{L}$). We treat $\sigma$ as the target ranking, and our objective is to update document embeddings so that the model’s ranking $\pi$ is transformed into the target ranking $\sigma$, while introducing minimal perturbations.

To formalize this process, we express the transformation from $\pi$ to $\sigma$ as a minimal sequence of adjacent swaps. 
Each swap imposes a specific constraint that, when enforced via optimization, yields local embedding adjustments sufficient to reverse the relative order of two adjacent documents in the original ranking.
The number of swaps required to convert $\pi$ into $\sigma$ is measured by the \textit{Kendall Tau distance} $K(\pi, \sigma)$, which counts the number of discordant pairs between the two permutations \cite{10.5555/644108.644113}:
\[
K(\pi, \sigma) = \left| \left\{ (i, j) : i < j,\ (\pi_i \prec \pi_j \text{ in } \pi \text{ and } \pi_i \succ \pi_j \text{ in } \sigma) \right\} \right|
\]
In this setting, the Kendall Tau distance provides a natural measure of the number of local ordering constraints that must be resolved to align the model’s ranking with the target one.

\paragraph{\textbf{Click-based Feedback via Counterfactual Modeling.}}
\label{sec:click}

To simulate interaction signals on the SERP, we use counterfactual models that estimate the likelihood of a user clicking a document based on its relevance and position. 
Building on previous work on counterfactual implicit feedback and online learning-to-rank systems \cite{ZhuangLiZuccon2022, JagermanOosterhuisDerijke2019, 10.1145/3269206.3271686, OvaisiAhsanEtAl2020, ZhuangQiaoZuccon2022, 10.1007/978-3-030-45439-5_28, OosterhuisDerijke2021, JoachimsSwaminathanSchnabel2017, WangBenderskyEtAl2016}, we simulate user-document interactions to generate synthetic click logs that reflect common biases in user behavior.

We model three main factors that influence user clicks: selection bias, position bias, and relevance bias.
\textit{Selection bias} arises because users rarely inspect all retrieved results. We assume that users examine results sequentially and stop at position $k'$ (i.e., only documents up to rank $k'$ are considered for clicks).
\textit{Position bias} captures the tendency of users to click more often on higher-ranked documents. It is modeled through an examination probability $\hat{p}_{e}(k) = \left(\frac{1}{k}\right)^\eta$, where $\eta$ controls the steepness of decay.
\textit{Relevance bias} reflects the likelihood of a click given a document’s relevance to the query. We define $\hat{p}_r(q, d)$ as the probability of clicking on document $d$ given its relevance with respect to query $q$. 

Following the literature, we define three user models to capture different user behaviors:

\begin{itemize}
\item \textbf{Perfect User (P):} Clicks deterministically on relevant documents. The click probability increases linearly with the document’s relevance score, ranging from 0 (non-relevant) to 1 (highly relevant).
\item \textbf{Noisy User (N):} Primarily clicks on relevant documents, but may also click on non-relevant ones with probability 0.2. Like Perfect User, click probabilities increase linearly with the relevance grade.
\item \textbf{Near-Random User (R):} Displays weak preference for relevant content, clicking non-relevant ones with probability 0.4 and highly relevant ones with 0.6. Intermediate relevance grades are interpolated linearly between these extremes.
\end{itemize}

\begin{table}[!t]
\centering
\caption{Simulated click probabilities for collections with four relevance levels: 0 = not relevant, 1 = partially relevant, 2 = relevant, 3 = highly relevant.}
\label{tbl:click_proba}
\begin{tabular}{r|cccc}
\toprule
 & \multicolumn{4}{c}{Document relevance score} \\
\multicolumn{1}{r|}{User model}     & \multicolumn{1}{c}{0} & \multicolumn{1}{c}{1} & \multicolumn{1}{c}{2} & \multicolumn{1}{c}{3} \\
\midrule
Perfect (P)          & 0.00 & 0.33 & 0.67 & 1.00 \\
Noisy (N)            & 0.20 & 0.40 & 0.60 & 0.80 \\
Near-Random (R)      & 0.40 & 0.47 & 0.53 & 0.60 \\
\bottomrule
\end{tabular}
\end{table}

\section{Experimental Settings}
\label{sec:settings}

Table~\ref{tbl:click_proba} summarizes the click probabilities used for each user model over a four-grade relevance scale. 
The final simulated click probability for a document $d$ shown at rank $i$ in response to query $q$ combines the relevance and position components:\begin{equation}
    p_c(q, d, i) = \hat{p}_r(q, d) \cdot \hat{p}_e(i).
\label{eq:clickP}
\end{equation}

In our setting, we set $k'=20$ and $\hat{p}_e(i) = 1$ for all positions, effectively removing position bias so that $p_c(q, d, i)$ depends only on the relevance-based component $\hat{p}_r(q, d)$.
For each query $q$, we assume users examine results sequentially and click documents according to Eq.~\ref{eq:clickP}. A click on document $d_r$ at rank $i>1$ is interpreted as a preference over the document $d_n$ at rank $i-1$. The resulting pair $(d_r,d_n)$ is added to $\mathcal{R}$, triggering an embedding surgery update.

This section outlines the experimental setup, including the datasets, dense retrieval models, evaluation metrics, and implementation details. The following research questions guide our design:

\textbf{RQ1 – Overall Effectiveness.}
\emph{How much does embedding surgery improve retrieval effectiveness across different relevance feedback sources, and which variant (symmetric, promotion, or demotion) is most effective? How stable is the method under noisy or conflicting feedback? Are the ranking improvements consistent across datasets and dense retrieval models, including out-of-domain settings? Do they effectively enforce the intended document ordering?}

\textbf{RQ2 – Impact on Latent Space Structure.}
\emph{What is the magnitude of the embedding updates? Does large-scale embedding surgery affect the structure of the latent representation space? How are the regions of the embedding space targeted by queries that share the same relevant documents affected?}

\textbf{RQ3 – Practicality and Scalability.}
\emph{Is embedding surgery practical for real-world deployment in large-scale retrieval systems? Can in-place embedding updates be applied to ANN indexes such as IVF or HNSW without degrading their structure or requiring costly re-indexing at scale?}

\textbf{RQ4 – Complementarity with Query Adaptation.}
\emph{Does embedding surgery complement query adaptation methods based on counterfactual user modeling? Specifically, does combining embedding surgery with CoRocchio yield additional gains over each method applied independently on dense retrieval benchmarks?}

\paragraph{\textbf{Datasets}}

We evaluate our embedding surgery framework on the following dense retrieval benchmarks:

\begin{itemize}

    \item \textbf{TREC DL 2019 (\dlXIX) and 2020 (\dlXX)}: 43 and 54 queries, respectively, with graded relevance judgments on a four-point scale over the MS MARCO passage collection ($\sim$8.8M passages) \cite{CraswellMitraEtAl2019,CraswellMitraEtAl2020}.
    
    \item \textbf{DL-Hard}: a challenging subset of queries from \dlXIX and \dlXX characterized by semantic ambiguity, lexical mismatch, and limited supervision \cite{DL_Hard}.
    
    \item \textbf{TREC \rbIV}: 249 queries (from the title field) over the TIPSTER corpus (disks 4–5, excluding congressional records) \cite{10.1145/1067268.1067272}. 
    Since the retrieval models used are fine-tuned on MS MARCO, this represents a zero-shot, out-of-domain setting.
    
    \item \textbf{TREC CAsT 2019}: 20 conversational search sessions ($\approx$10 turns each) with graded relevance judgments over 38.5M passages from TREC CAR, WaPo, and MS MARCO \cite{dalton_cast_2019,10.1145/3578519}. We use rewritten utterances to study how surgery on initial queries affects subsequent context-dependent queries.
    
    \item \textbf{QSharedRel}: A subset of 346 queries extracted from MS MARCO Dev in which pairs of queries share at least one relevant passage. Among these, 169 are designated as focal queries, for which embedding surgery is applied. We then evaluate its effect on the corresponding related queries to assess whether ranking corrections generalize across semantically related information needs.
    
    \item \textbf{MS MARCO Dev}: $\approx$56k judged queries used to analyze large-scale embedding updates and their impact on the latent space and ANN indexing.

\end{itemize}

\paragraph{\textbf{Dense retrieval models}}
We assess the generalizability of embedding surgery across different dense semantic representations.

\begin{itemize}
    \item \textbf{Contriever}~\cite{IzacardCaronEtAl2021}: A dense retriever trained with contrastive learning on unlabeled web data and MS MARCO.
    
    \item \textbf{TAS-B}~\cite{HofstatterLinEtAl2021}: A BERT-based dual encoder fine-tuned on MS MARCO and widely used for passage ranking.

    \item \textbf{E5 (Multilingual)}~\cite{DBLP:conf/acl/WangYHYMW24}: A multilingual sentence embedding model trained on diverse tasks including retrieval, classification, and summarization. 

    \item \textbf{Snowflake-Arctic Embed}~\cite{yu2024arcticembed20multilingualretrieval}: A general-purpose embedding model trained by Snowflake for high-coverage similarity and semantic search applications.
\end{itemize}

All models are used in a zero-shot setting, without task-specific fine-tuning beyond their publicly released checkpoints.

\paragraph{\textbf{Evaluation metrics}}

We evaluate the results by using standard retrieval quality metrics and additional measures that capture the magnitude, impact, and cost of embedding modifications. 

Specifically, we evaluate search quality before and after surgery using
\textbf{nDCG@k} (Normalized Discounted Cumulative Gain), which measures the quality of the top-k retrieved documents, taking into account both graded relevance and ranking position. 
\textbf{RR} (Reciprocal Rank) or \textbf{MRR} (Mean Reciprocal Rank)  are instead used for datasets containing a single relevant item per query, such as MS MARCO Dev and QSharedRel.
Surgery accuracy is assessed using \textbf{Kendall Tau distance} (KD), which measures the number of pairwise inversions between the pre-surgery and target rankings, and \textbf{Swap Accuracy}, defined as the percentage of document pairs correctly ordered after surgery.

\paragraph{\textbf{Implementation details}}

We implement embedding surgery as a modular Python library compatible with standard dense retrieval pipelines built on FAISS. For each query, we retrieve the top-\(20\) results and apply embedding surgery as discussed earlier. Qualitatively similar trends are observed across other cutoffs. We use FAISS Flat indexes for all experiments addressing RQ1, RQ2, and RQ4, while, for RQ3, we additionally consider IVF and HNSW ANN indexes. 
For IVF, we set the number $c$ of clusters to $\sqrt{|\mathcal{D}|}$, and nprobe=$c/10$ \cite{jegou2010product,babenko2014inverted}. For HNSW, we set the graph connectivity degree $M$ to $32$, efConstruction to $400$, and efSearch to $256$, corresponding to a high-recall configuration \cite{8594636}.
The embedding optimization problem is solved using the \texttt{cvxpy} library \cite{diamond2016cvxpy}. We evaluate the computational cost of the optimization on a single core of an AMD Rome 7742 CPU running at 2.25 GHz. To limit perturbations in the latent space, the margin parameter \(\epsilon\) in Eq.~\ref{eq:surgery} is set to the minimum between the score difference within each constrained document pair and a fixed threshold of \(0.01\). 
Embeddings are not re-normalized after applying $\Delta \mathbf{d}_i$; scores are computed as raw inner products for all encoders, and the resulting relative norm change is bounded by the small update magnitudes we observe (§\ref{sec:RQ2}). We empirically observed that re-normalizing after the update degrades effectiveness, confirming this design choice.
Optimization is performed independently for each query, and the resulting updates are applied immediately to the index. Only the embeddings of documents involved in the constraints are modified, while all other document embeddings remain unchanged. Because the constraints generated by all three feedback sources are derived from a single target ranking, the induced preference relation is always acyclic and Eq. \ref{eq:surgery} is guaranteed feasible. For generating LLM-based relevance feedback, we employ the Qwen\footnote{\url{https://huggingface.co/Qwen/Qwen3-Reranker-8B}} and multilingual BGE\footnote{\url{https://huggingface.co/BAAI/bge-reranker-v2-gemma}} rerankers. Finally, retrieval performance is evaluated using \texttt{pytrec\_eval}.

The full implementation is publicly available, enabling easy reproduction of all experiments\footnote{\url{https://github.com/maddalena-amendola/Embedding-Surgery}}.

\section{Results and Analysis}
\label{sec:experiments} 

\subsection{RQ1 - Overall Effectiveness}
\label{sec:RQ1}

\begin{figure*}[t]
    \centering
    \includegraphics[width=0.85\textwidth, trim={0 2mm 0 2mm},clip]{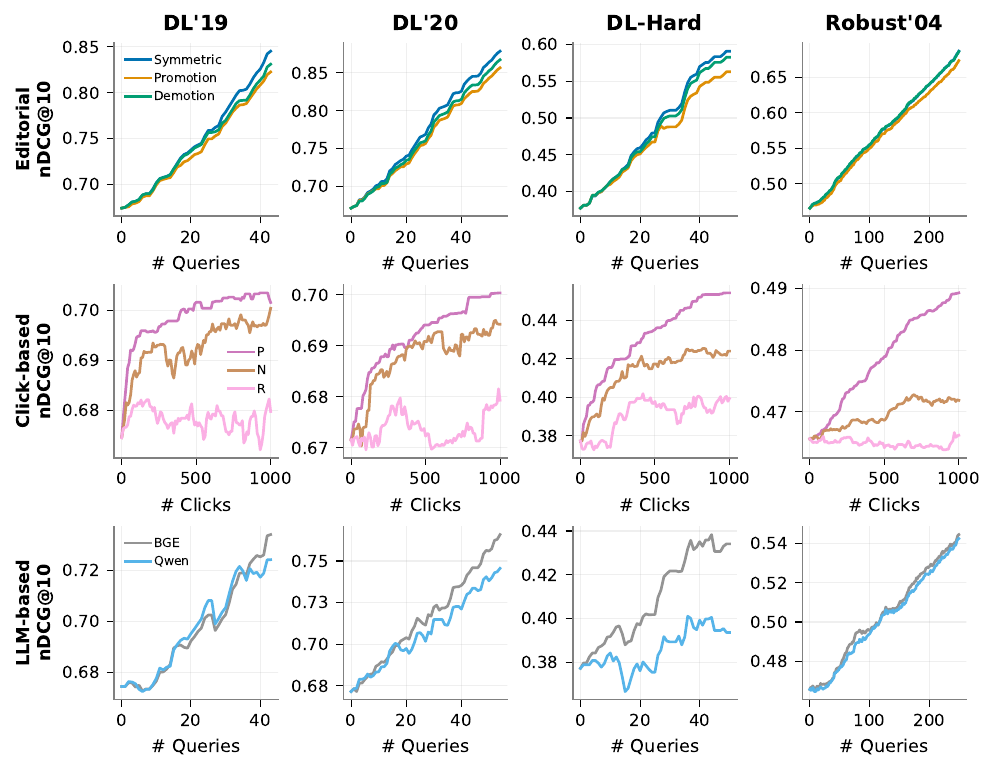}
    \caption{nDCG@10 results on \dlXIX, \dlXX, DL-Hard, and \rbIV using Contriever embeddings across editorial (top row), click-based (middle row), and LLM-based (bottom row) feedback scenarios. In the click-based setting, curves correspond to different user models: perfect (P), noisy (N), and near-random (R). }
    \label{fig:DL_robust}
\end{figure*}

To answer RQ1, we evaluate the overall effectiveness of embedding surgery across different datasets, dense encoding models, and supervision sources. Figure~\ref{fig:DL_robust} reports the average nDCG@10 obtained with the Contriever model on the \dlXIX, \dlXX, DL-Hard, and \rbIV datasets. 
The plots report the average nDCG@10 as embedding surgery is progressively applied to all the queries in each dataset.

The plots in the first row refer to the editorial feedback scenario, where ranking corrections via surgery on the top-20 retrieved results are guided by the gold standard in an oracular setting. 
As described in Section~\ref{sec:methodology}, we investigated three variants of the embedding surgery formulation (i.e., symmetric, demotion, and promotion) that modify both positive and negative embeddings, only negative embeddings, or only positive embeddings, respectively.
Looking at the nDCG@10 curves, it is evident that the symmetric surgery method is preferable to the other two methods across all datasets, except \rbIV, where its performance is very close to that of the demotion setting. The symmetric variant results, in fact, in the largest effectiveness gains, with a monotonic increase in nDCG@10 from 0.674, 0.672, 0.377, and 0.466 pre-surgery to 0.845, 0.878, 0.591, and 0.687 for \dlXIX, \dlXX, DL-Hard, and \rbIV, respectively. Consequently, all subsequent experiments will rely on the symmetric surgery method. 

The plots in the middle row of Figure~\ref{fig:DL_robust} regard the click-based feedback scenario and show how nDCG@10 varies with the number of simulated user clicks for the perfect (P), noisy (N), and near-random (R) counterfactual user models defined in Section~\ref{sec:click}. The first observation from these plots is that embedding surgery consistently enhances ranking performance across all four datasets. As expected, the setting with the perfect user model achieves the largest effectiveness gains, reaching 0.702, 0.700, 0.454, and 0.486 after $1$,$000$ clicks for \dlXIX, \dlXX, DL-Hard, and \rbIV, respectively.  
To address the stability sub-question of embedding surgery under noisy or conflicting feedback, we observe that embedding surgery based on implicit click feedback is robust to variations in user behavior. Even when click signals are conflicting,  performance remains stable: under the noisy user model (N), nDCG@10 improvements are consistent and very close to those achieved with a perfect user on \dlXIX and \dlXX. Near-random click behavior (R), in contrast, yields limited gains but does not degrade the original retrieval effectiveness, except for \rbIV, where we observe a marginal, not statistically significant decrease.

Finally, the plots in the bottom row of Figure~\ref{fig:DL_robust} refer to the LLM-based feedback scenario, where BGE and Qwen are used to rerank top-20 results retrieved with Contriever, and the resulting ranking is used to perform surgery as detailed in Section~\ref{sec:LLM-based}. Also in this scenario, where again we apply surgery incrementally to each query of the benchmark, we observe consistent improvements in ranking performance across all four datasets, even if the ranking of some queries is penalized by the low quality of the LLM-based reranking, as indicated by the local drops in average performance visible in the nDCG@10 curves. For DL-Hard, the Qwen reranking model exhibits relatively poor performance, and consequently, surgery guided by this reranking yields only marginal benefits.

Interestingly, across both click-based and editorial feedback sources, the largest performance gains from surgery are observed in the DL-Hard dataset, which contains difficult queries, highlighting the potential of our technique to address cases where ranking models (and even LLMs) underperform.

\begin{table*}[!t]
    \caption{Retrieval effectiveness (nDCG@10), average KD, and swap accuracy (\%) for click-based (perfect user), LLM-based, and editorial feedback across different dense models and datasets considering the symmetric surgery approach. }
    \label{tab:models}
    \centering
    \small
    
    \begin{tabular}{llc|rcr|rcr|rr}
    \toprule
      &  & \textbf{Baseline} & \multicolumn{3}{c}{\textbf{Editorial}} &  \multicolumn{3}{|c|}{\textbf{LLM-based (BGE)}} &  \multicolumn{2}{c}{\textbf{Click-based (perfect user)}}\\

    & & \textbf{nDCG@10} & \textbf{nDCG@10 ($\Delta$\%)} & \textbf{KD} & \textbf{Swap (\%)} & \textbf{nDCG@10 ($\Delta$\%) }& \textbf{KD} & \textbf{Swap (\%)} &  \textbf{nDCG@10 ($\Delta$\%)} & \textbf{Swap (\%)}\\
    \midrule
      \multirow{ 4}{*}{\dlXIX}  & Contriever    & 0.674 & 0.845 (+25.37) & 37.98 & 100.00 & 0.734 (+8.90) & 55.58 & 99.64 & 0.702 (+4.15) & 100.00  \\
         &  TAS-B                               & 0.717 & 0.876 (+22.18) & 36.33 & 100.00 & 0.749 (+4.46) & 59.79 & 99.37 & 0.741 (+3.35) & 100.00  \\
         &  E5                                  & 0.703 & 0.866 (+23.19) & 37.45 & 100.00 & 0.753 (+7.11) & 54.35 & 100.00 & 0.730 (+3.84) & 100.00 \\
         &  Snowflake                           & 0.725 & 0.883 (+21.79) & 34.93 & 100.00 & 0.751 (+3.59) & 50.74 & 100.00 & 0.750 (+3.45) & 100.00 \\
    \midrule
      \multirow{ 4}{*}{\dlXX}     & Contriever  & 0.672 & 0.878 (+30.65) & 38.37 & 99.05 & 0.766 (+13.99) & 56.39 & 98.54 & 0.700 (+4.17) & 100.00 \\
         &  TAS-B                               & 0.684 & 0.868 (+26.90) & 35.07 & 99.68 & 0.755 (+10.38) & 54.43 & 98.75 & 0.711 (+3.95) & 100.00  \\
         &  E5                                  & 0.698 & 0.863 (+23.64) & 35.31 & 99.82 & 0.765 (+9.60) & 51.22 & 99.91 & 0.725 (+3.87) & 100.00 \\
         &  Snowflake                           & 0.733 & 0.878 (+19.78) & 31.45 & 100.00 & 0.767 (+4.64) & 50.87 & 100.00 & 0.750 (+2.32) & 100.00  \\
    \midrule
      \multirow{ 4}{*}{DL-Hard}   & Contriever  & 0.377 & 0.591 (+56.76) & 28.64 & 99.78 & 0.434 (+15.12) & 63.96 & 98.69 & 0.454 (+20.42) & 100.00  \\
         &  TAS-B                               & 0.376 & 0.604 (+60.64) & 28.93 & 99.81 & 0.419 (+11.44) & 66.66 & 99.17 & 0.469 (+24.73) & 85.71  \\
         &  E5                                  & 0.368 & 0.573 (+55.71) & 28.26 & 100.00 & 0.420 (+14.13) & 61.50 & 99.27 & 0.460 (+25.00) & 100.00 \\
         &  Snowflake                           & 0.381 & 0.612 (+60.63) & 30.59 & 100.00 & 0.432 (+13.39) & 58.14 & 99.40 & 0.480 (+25.98) & 100.00 \\
    \midrule 
      \multirow{ 4}{*}{\rbIV}    & Contriever   & 0.466 & 0.687 (+47.42) & 35.35 & 99.95 & 0.545 (+16.95) & 71.63 & 99.72 & 0.487 (+4.51) & 100.00 \\
         &  TAS-B                               & 0.447 & 0.644 (+44.07) & 33.15 & 99.90 & 0.505 (+12.98) & 70.69 & 99.83 & 0.470 (+5.15) & 100.00 \\
         &  E5                                  & 0.395 & 0.619 (+56.71) & 37.30 & 99.58 & 0.492 (+24.56) & 73.90 & 99.63 & 0.428 (+8.35) & 100.00 \\
         &  Snowflake                           & 0.466 & 0.683 (+46.57) & 33.54 & 100.00 & 0.537 (+15.24) & 68.00 & 100.00 & 0.490 (+5.15) & 100.00 \\
    
    \bottomrule
    \end{tabular}
\end{table*}

Table~\ref{tab:models} reports the results addressing the last two sub-questions of RQ1, namely the generalizability of improvements across models and across in-domain and out-of-domain datasets when applying the symmetric surgery method.
For each dataset and model, the Baseline column shows the nDCG@10 before any modification, serving as a common baseline reference across all feedback conditions. For each feedback method, the nDCG@10 ($\Delta$\%) column reports the post-surgery nDCG@10 and the relative improvement over the pre-surgery nDCG@10. 

We observe consistent performance gains across all models, datasets, and feedback strategies, confirming the general effectiveness of embedding surgery (all improvements are statistically significant under a paired two-sided t-test with p < 0.05, except for Snowflake in the LLM-based setting on \dlXIX). As expected, Editorial feedback, which relies on human-assessed ground-truth labels, yields the largest improvements, reaching up to +60.64\% for TAS-B on DL-Hard and never dropping under +19.78\% (Snowflake, \dlXX). The LLM-based setting, which uses BGE to produce rerankings, also achieves stable gains, with the smallest improvement of +3.59\% (Snowflake, \dlXIX) and the largest of +24.56\% (E5, \rbIV). Finally, the Click-based feedback scenario, which here reflects 1,000 interactions with perfect users, 
improves consistently, with the smallest gain of +2.32\% for Snowflake on \dlXX and the highest of +25.98\% with Snowflake on DL-Hard. 
Notably, improvements in the out-of-domain \rbIV setting are comparable to those observed in-domain, indicating that the method generalizes beyond the training distribution and remains effective under distribution shift.
Overall, across all models, benchmarks, and feedback scenarios, embedding surgery consistently improves nDCG@10 scores, with the largest gains observed on DL-Hard and \rbIV. These results confirm the robustness of the optimization process, the advantage of fixing embedding representations per query, and the method’s effectiveness on particularly challenging queries.

To address the last sub-question, the KD and Swap (\%) columns in the table report (i) the average number of pairwise ordering constraints per query (i.e., the KD) required to align the rankings, and (ii) the average percentage of ordering constraints from Eq.~\ref{eq:surgery} that are satisfied during retrieval after surgery, respectively.
The KD column is omitted for the click-based setting where at most one constraint is generated per interaction, since each user click identifies a document deemed more relevant than the previously examined one.

We observe that LLM-based feedback produces the longest constraint lists, reflecting the greater variability and lower reliability of LLM-generated rankings compared with rankings derived from editorial judgments. A larger number of constraints naturally increases task complexity, resulting in slightly lower swap accuracy. By contrast, rankings derived from editorial annotations yield shorter constraint lists, averaging 28.26--38.37 constraints per query.

The click-based setting, which involves only a single document pair per update, achieves perfect or near-perfect swap accuracy in most cases, except for TAS-B on DL-Hard. As expected, the LLM-based and editorial settings yield slightly lower yet still very high swap accuracies (98.54\% and 99.05\%, respectively), confirming the robustness of the optimizer under noisier supervision and its ability to consistently enforce the desired ranking order after surgery.

\begin{figure}[!t]
    \centering
    \includegraphics[width=0.85\columnwidth, trim={0 3mm 0 3mm}, clip]{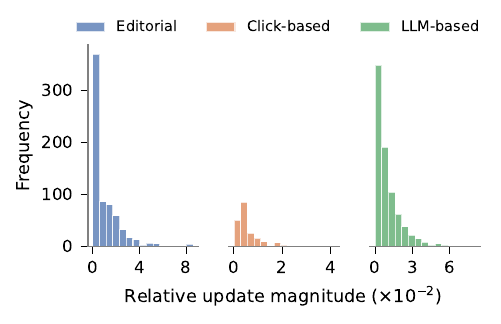}
    \caption{Distribution of embedding update magnitudes (normalized $\ell_2$ norms, Contriever, \dlXIX dataset).}
    \label{fig:norm}
\end{figure}

\subsection{RQ2 - Impact on Latent Space Structure}
\label{sec:RQ2}
To assess the robustness of the embedding surgery method and its impact on the structure of the latent space, we first analyze the magnitude of the perturbations it introduces into the embedding space. Figure~\ref{fig:norm} reports the distribution of normalized $\ell_2$ norms of the updates performed, measuring the average magnitude of $|\Delta \mathbf{d}_i|_2$ across all modified Contriever embeddings under the editorial, click-based, and LLM-based (BGE) feedback settings using  \dlXIX.

Across all feedback types, the update magnitudes remain small (i.e., typically below 0.05), indicating that embedding surgery induces minimal, localized modifications to document representations. The distributions are right-skewed and approximately log-normal, with a sharp concentration near zero and only a few larger corrections. Editorial and LLM-based feedback exhibit slightly heavier tails, reflecting the higher number of ranking constraints per query compared to the single-pair updates of the click-based scenario. Similar patterns are observed for the other models and settings.

Next, we further examine whether embedding surgery improves rankings without unintended side effects. First, we investigate whether large-scale embedding surgery affects the structure and stability of the latent space. To this end, we apply editorial-guided surgery to the Contriever MS MARCO passage embeddings using all 56k  queries from the MS MARCO Dev judged set, and evaluate whether retrieval effectiveness is preserved on downstream benchmarks, namely the TREC DL and DL-Hard test sets. Concretely, we compare retrieval effectiveness on the same FLAT index before and after applying large-scale surgery, while keeping the retrieval pipeline unchanged, to isolate the impact of the embedding updates on ranking quality.  
The results show that large-scale embedding surgery has a negligible impact on the latent space's structure. As expected, nDCG@10 on the judged MS MARCO Dev set increases substantially, from 0.407 to 0.716. In contrast, performance on the held-out DL benchmarks remains essentially stable: on \dlXIX, nDCG@10 slightly decreases from 0.674 to 0.669, while on \dlXX it changes marginally from 0.672 to 0.670. A similar pattern is observed for Recall@100, which remains unchanged at 0.530 on \dlXIX and decreases only marginally from 0.586 to 0.585 on \dlXX. On DL-Hard, performance also remains stable, with nDCG@10 changing from 0.377 to 0.373 and Recall@100 increasing slightly from 0.492 to 0.494. None of these differences is statistically significant according to a paired t-test ($p > 0.05$).
Overall, these negligible variations indicate that, despite applying embedding surgery at scale using all 56k judged queries, the updates do not introduce measurable collateral degradation in retrieval quality across the evaluation benchmarks. This stability supports the conclusion that the induced modifications are small and localized, preserving the overall structure of the latent space while introducing only limited, targeted adjustments.

\begin{figure*}[!t]
    \centering
    \includegraphics[width=0.9\textwidth, trim={0 1mm 0mm 1mm},clip]{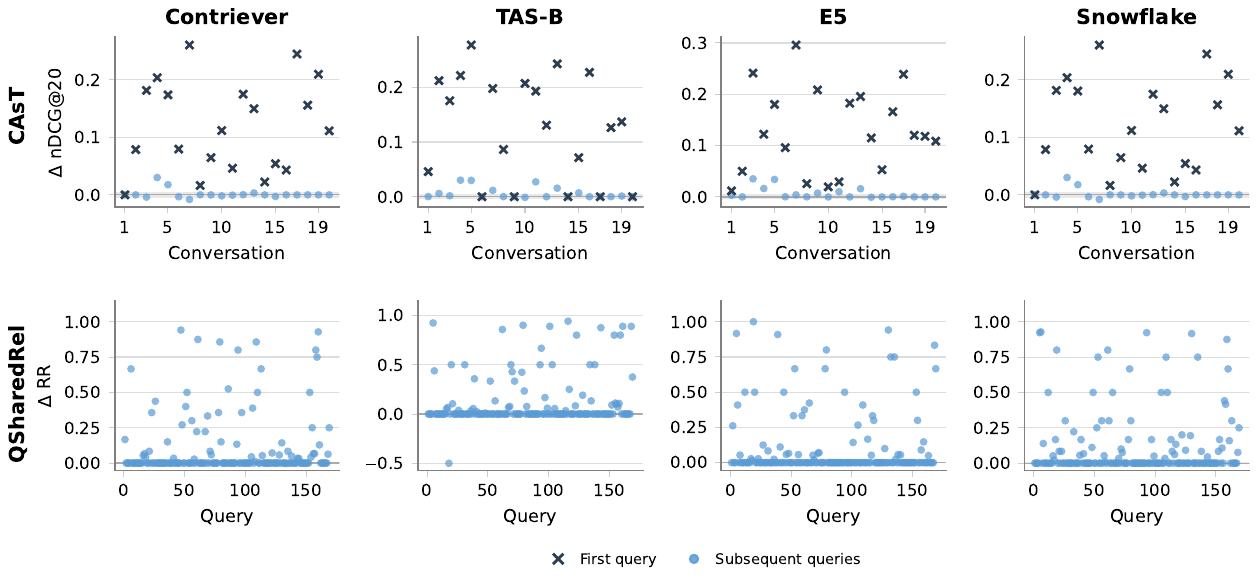}
    \caption{Changes in nDCG@20 and RR ($\Delta$ nDCG@20 and $\Delta$ RR)  observed on related queries after applying surgery to the first query of each conversation in CAsT, and to the first query of each group in QSharedRel.}
    \label{fig:cast-dev}
\end{figure*}

To answer the last sub-question of RQ2, we study how local corrections propagate to semantically related queries that may share relevant documents by conducting two complementary experiments on the CAsT and QSharedRel datasets. In CAsT, multiple utterances within the same conversation often retrieve overlapping documents and share relevance annotations~\cite{dalton_cast_2019,10.1145/3578519,10.1145/3404835.3463090}. In contrast, QSharedRel consists of pairs of distinct MS MARCO Dev judged queries that share at least one relevant passage, indicating overlap in their information needs.
In the CAsT experiment, we apply editorial-guided embedding surgery only to the first query of each conversation and measure the $\Delta$ nDCG@20 for subsequent utterances. This setup verifies that local interventions do not disrupt the latent space, ensuring that modified embeddings preserve retrieval consistency for related queries in a conversational setting. For QSharedRel, we perform surgery on one query in each group and evaluate the $\Delta$ RR for the other query (typically a single paired query) sharing the same relevant passage. This experiment tests whether improvements to one query transfer to another that addresses the same underlying information need. The results are summarized in Figure~\ref{fig:cast-dev}, where rows correspond to datasets and columns to embedding models.

Examining the first row of Figure~\ref{fig:cast-dev} (CAsT), embedding surgery consistently improves the nDCG@20 of the first query in each conversation (black crosses), as all markers lie on the positive axis. In contrast, its effect on subsequent context-dependent queries (blue dots) is minimal: slight improvements in some cases, negligible drops in others, and often no change. This suggests that embedding surgery performs precise, localized adjustments that enhance the target query without perturbing the surrounding semantic region.

To further support this observation, we compute the overlap between the top-20 results before and after surgery. The intersection remains consistently high, near 20 and always above 19 documents, confirming that retrieval rankings are largely preserved.

For QSharedRel (second row of Figure~\ref{fig:cast-dev}), embedding surgery has a positive or neutral effect ($\Delta$ RR $\ge 0$) on paired queries in almost all cases. This result suggests that improvements applied to one query can transfer to semantically related queries with overlapping relevant documents. The results indicate that the modified embeddings better capture shared relevance signals, reinforcing rather than distorting the local semantic structure.
For example, with Contriever, performance improves in 54 cases and decreases slightly in only one case ($-0.003$ RR). Among the 114 unchanged cases, 91 have already achieved a reciprocal rank of 1 and cannot improve further. A notable drop ($-0.5$ RR) occurs in only one case with TAS-B, where the relevant passage shifts from rank 1 to 2. Such cases are extremely rare, confirming that embedding surgery typically benefits nearby queries while preserving overall stability.

\subsection{RQ3 - Practicality and Scalability}
\label{sec:RQ3}
To address RQ3 on practicality and scalability, we first measure the computational cost of solving the constrained optimization problem with click-based feedback (a single ranking preference constraint) and editorial feedback (a batch of constraints). Experiments are conducted using Contriever on \dlXIX. For the click-based setup, we measured the time required to perform 100 swaps, whereas the editorial setup relies on all \dlXIX test queries. 
Embedding surgery incurs low computational overhead: embedding updates take, on average, $7.9 \times 10^{-3}$ $(\pm 6.18 \times 10^{-5})$ seconds per constraint in the click-based setting, and $3.6 \times 10^{-3}$ $(\pm 5.68 \times 10^{-5})$ seconds per constraint in the batch editorial setting, highlighting the efficiency of jointly updating multiple documents. Overall, these latencies are modest and do not hinder practical deployment, as the optimization can be overlapped with other computations of the retrieval pipeline.

Beyond the cost of solving the optimization problem, embedding surgery also requires updating modified vectors in the index, whose cost may vary depending on the index type and configuration \cite{10.1007/s00778-024-00864-x}. One option is to perform an upsert, i.e., removing the original vector and reinserting the updated one. A more efficient alternative, particularly when updates are small (as in our case; see Section~\ref{sec:RQ2}), is to overwrite embeddings in place, which incurs negligible overhead.

An open question is whether directly overwriting embeddings affects ANN index performance, potentially requiring periodic background consolidation to preserve index integrity. To investigate this issue, we consider two widely used ANN index structures: graph-based HNSW and cluster-based IVF indexes. Specifically, we evaluate: (i) whether in-place embedding updates in HNSW indexes (i.e., without modifying graph edges) affect retrieval effectiveness or graph navigation quality; and (ii) the extent to which embedding updates trigger reassignment across IVF clusters, where a low reassignment rate would suggest that updates can be handled without modifying the index structure.

For both HNSW and IVF, we build indexes over MS MARCO passage embeddings produced by Contriever and evaluate baseline retrieval effectiveness on \dlXIX, \dlXX, and DL-Hard. Following the setup used in RQ2, we then apply large-scale editorial-guided surgery using all 56k judged MS MARCO Dev queries, update embeddings in place, and re-evaluate retrieval performance on the same benchmarks.
As a result, 185,006 embeddings in the HNSW index are overwritten without modifying graph connectivity. For IVF, surgery modifies 183,106 embeddings, of which only 691 ($\approx$0.38\%) would change cluster assignment after the update (without reassignment).

\begin{table}[!t]
\centering
\caption{
Impact of large-scale surgery on ANN search (Contriever)
using the 56k MS MARCO Dev judged queries (DEV).
}
\label{tab:large_scale_surgery}
\begin{tabular}{llcccc}
\toprule
& & \multicolumn{2}{c}{\textbf{HNSW}}  & \multicolumn{2}{c}{\textbf{IVF}} \\ \cmidrule(lr){3-4} \cmidrule(lr){5-6} \textbf{Dataset} & \textbf{Metric} & \textbf{Before} & \textbf{After} & \textbf{Before} & \textbf{After} \\\midrule
DEV                      & MRR & 0.348 & 0.714 & 0.348 & 0.709 \\ \midrule
\multirow{2}{*}{\dlXIX}  & nDCG@10 & 0.670 & 0.664 & 0.652 & 0.647 \\ 
                         & Recall@100 & 0.530 & 0.530 & 0.520 & 0.520 \\ \midrule
\multirow{2}{*}{\dlXX}   & nDCG@10 & 0.668 & 0.667 & 0.667 & 0.665 \\ 
                         & Recall@100 & 0.579 & 0.578 & 0.566 & 0.566 \\ \midrule 
\multirow{2}{*}{DL-Hard} & nDCG@10 & 0.377 & 0.373 & 0.377 & 0.373 \\                                                                   & Recall@100 & 0.492 & 0.494 & 0.479 & 0.483 \\ \bottomrule
\end{tabular}
\end{table}

Table~\ref{tab:large_scale_surgery} reports the results assessing ANN index effectiveness before and after embedding overwrites on the MS MARCO Dev and DL benchmarks. As expected, surgery substantially improves effectiveness on the judged MS MARCO Dev queries, where MRR increases from 0.348 to 0.714 and 0.709 for HNSW and IVF, respectively.
The effect of in-place updates on the HNSW index is minimal across all DL benchmarks. On \dlXIX, nDCG@10 decreases slightly from 0.670 to 0.664, while Recall@100 remains unchanged at 0.530. Almost no variation is observed on \dlXX (0.668 to 0.667 in nDCG@10 and 0.579 to 0.578 in Recall@100) and DL-Hard (0.377 to 0.373 in nDCG@10, with a slight increase in Recall@100 from 0.492 to 0.494). These small variations indicate that HNSW is robust to in-place embedding updates, and that modifying vectors without adjusting graph connectivity does not significantly affect navigation quality.
Similar observations hold for IVF. On \dlXIX, nDCG@10 decreases slightly from 0.652 to 0.647, with Recall@100 unchanged at 0.520. On \dlXX and DL-Hard, the impact on nDCG@10 is again negligible (0.667 to 0.665 on \dlXX and 0.377 to 0.373 on DL-Hard), with a slight increase in Recall@100 for DL-Hard (0.479 to 0.483). 

Overall, no consistent degradation in retrieval effectiveness is observed (none of the small differences in performance are statistically significant according to a paired t-test with $p > 0.05$), indicating that surgery can be applied through in-place embedding overwrites to both graph-based and cluster-based indexes without requiring index reconstruction. This stability stems from the small, controlled nature of the updates, which introduce only localized changes while preserving the structure of the representation space. Importantly, this makes embedding surgery practical for ANN-based retrieval systems without incurring costly maintenance.

\begin{table}[!t]
\centering
\caption{nDCG@10 on \dlXIX using Contriever with CoRocchio, embedding surgery, and their combination.}
\label{tab:corocchio_surgery}

\begin{tabular}{lcccc}
\toprule
\textbf{User Model}
& \textbf{Baseline} & \textbf{+ CoR.} & \textbf{+ Surg.} & \textbf{+ CoR.+Surg.} \\ \midrule
Perfect     & 0.674 & 0.799 & 0.702 & \textbf{0.809} \\
Noisy       & 0.674 & 0.699 & 0.700 & \textbf{0.720} \\
Near-random & 0.674 & 0.632 & \textbf{0.680} & 0.645 \\ \bottomrule
\end{tabular}
\end{table}

{\subsection{RQ4 - Complementarity with Query Adaptation Methods}
\label{sec:corocchio}
To answer the last RQ, we consider CoRocchio \cite{ZhuangLiZuccon2022} as a representative query adaptation method and evaluate embedding surgery and CoRocchio independently and in combination.
Specifically, we use Contriever embeddings and the original implementation of CoRocchio from \cite{ZhuangLiZuccon2022}, which simulates $1{,}000$ interactions per query on \dlXIX under the three user models described in Section~\ref{sec:click}. Query representations are then updated based on the collected signals, and effectiveness is measured using nDCG@10. In contrast, embedding surgery is applied in an event-driven manner using $1{,}000$ interactions in total across all test queries. For the combined CoRocchio+Surgery setting, we first update query representations with CoRocchio and then apply embedding surgery using the same additional global budget of $1{,}000$ interactions. Since \dlXIX contains 43 queries, the combined setting uses only a marginally larger feedback budget than CoRocchio alone, while allowing us to test whether document-side corrections provide complementary gains to query-side adaptation.
The results are reported in Table~\ref{tab:corocchio_surgery}. Under a perfect user model, CoRocchio improves nDCG@10 from 0.674 to 0.799, while embedding surgery alone reaches 0.702; their combination further increases effectiveness to 0.809, indicating clear complementarity. A similar trend holds under noisy feedback, where the combined approach (0.720) outperforms both CoRocchio (0.699) and surgery (0.700). However, under near-random feedback, CoRocchio degrades performance (0.632), whereas embedding surgery remains stable (0.680), and the combination (0.645) inherits part of CoRocchio’s degradation.
This behavior reflects the different feedback assumptions of the two methods. CoRocchio adapts query representations by aggregating multiple clicks on relevant documents, thereby benefiting from richer, more reliable signals. However, its updates remain query-specific and do not propagate to semantically related queries. In contrast, embedding surgery modifies document embeddings directly within the shared representation space. As shown in previous experiments, these updates can transfer improvements to semantically related queries with overlapping relevance patterns. Under counterfactual modeling, however, surgery operates on single interactions, which limits its ability (under perfect feedback) to propagate relevance beyond top-ranked documents but also makes it less sensitive to noisy or conflicting signals. Overall, embedding surgery complements query adaptation when feedback is informative, while offering greater robustness and broader generalization under unreliable user behavior.

\section{Conclusion}
\label{sec:conclusion}

This work introduces embedding surgery, a novel framework for real-time ranking correction through lightweight, feedback-driven updates of document representations in the latent space. This approach enables fine-grained modifications at query time, guided by relevance signals from user interactions or explicit human/LLM assessments. It is particularly suited for correcting known retrieval issues or enabling editorial interventions, such as improving fairness, mitigating biases, or enforcing business rules. 
By formulating the problem as a constrained convex optimization, we ensure minimal and controlled changes to the embedding space.

Experiments on seven IR benchmarks show that embedding surgery consistently improves ranking effectiveness with negligible computational overhead and generalizes well under distribution shift. By incorporating updates into the index, embedding surgery can also propagate these improvements to semantically related queries, a valuable property for a post-hoc correction method. We also show that embedding surgery complements query adaptation methods such as CoRocchio, yielding additional gains under reliable feedback while providing greater robustness under noisy user interactions.
Moreover, large-scale experiments using all 56k judged queries from MS MARCO Dev further demonstrate that embedding surgery preserves the effectiveness of the collection representation on held-out downstream benchmarks, confirming that these localized updates do not disrupt the global structure of the latent space. Moreover, experiments with scalable ANN indexes show that embeddings can be updated efficiently via in-place overwriting with negligible impact on the index structure and navigation, thereby avoiding costly index reorganization.

This work opens a new direction for adaptive vector search, enabling fine-grained, feedback-driven ranking correction without retraining or costly index reconstruction. 
Future work will explore the integration of embedding surgery with sparse learned representations, systematically analyze sensitivity to the margin $\epsilon$ and surgery cutoff, investigate constraint-relaxation strategies for inconsistent feedback, evaluate more realistic click models and real-world interaction logs, and consider additional baselines, including online ranking models.

\section*{GenAI Usage Disclosure}
The authors used GenAI tools for language editing and occasional support with code development and debugging. Any suggestions or code produced by these tools were carefully reviewed, revised, and validated by the authors. GenAI tools were not used to generate experimental results, data, citations, or the core scientific contributions of the work. The authors remain fully responsible for the content and results presented in this paper.

\balance
\bibliographystyle{ACM-Reference-Format}
\bibliography{biblio}
\end{document}